# Are There Echo Chambers in the US News Ecosystem? Evidence From Twitter/X

Wen Yang

The Hong Kong University of Science and Technology (Guangzhou)

wyangbl@connect.ust.hk

**Abstract**

This study investigates echo chambers in social networks through an analysis of Twitter news accounts. Utilizing bias labels from the AllSides website, we construct a dataset representing six dimensions of news bias. Through manual extraction of follower/following relationships, we analyze interactions among 65 active Twitter news accounts. Despite the relatively small size of the network node data utilized, results reveal distinct clustering patterns indicative of echo chambers, with limited interaction between conflicting ideologies. This study underscores the potential impact of bias on information dissemination and democratic expression. These findings offer valuable insights into the dynamics of echo chambers in contemporary social media environments.

## 1. Introduction

Echo chambers have become a focal point in social network research. Internal factors contributing to the formation of echo chambers include human cognitive and psychological biases, such as confirmation bias, as well as tendencies in social interactions to maintain self-esteem and seek like-minded individuals [1]. External factors involve the reinforcement of similar opinions and the rejection of dissenting views through communication channels and environmental characteristics [2]. Notably, social media algorithms tend to amplify information aligning with users' past preferences, exemplified by Eli Pariser's renowned study on algorithmic filter bubbles [3]. His research revealed that Facebook's recommendation algorithm caters to users with Democratic/Republican tendencies, inundating their information streams with content reflecting similar political stances and ideologies. This undoubtedly exacerbates societal divisions and political polarization, hindering the expansion of perspectives and diverse communication.

Studies indicate that the political landscape, particularly in systems employing referendums/elections, has been significantly influenced by echo chamber effects since 2016 [4], [5]. Scandals such as the Cambridge Analytica controversy in 2018 [6], [7] and the 2020

Netflix documentary "The Social Dilemma"[1] exposed how social media giants intertwine with political campaigns to manipulate users' minds, cognition, and behavior. However, the existence of echo chambers in social networks is a point of contention among researchers. Some studies, employing various metrics or experimental recruitment, attempt to disprove the significant presence of echo chamber effects, dismissing them as exaggerations [8].

Therefore, investigating echo chamber in social networks, obtaining evidence from real social data, holds profound implications for shaping societal awareness, enhancing regulators' attention, promoting updates to social network architectures, and even driving humanitarian capital and sustainable economic development [9]. This paper employs data analysis of news accounts on Twitter to explore evidence of echo chamber effects through underlying ideological bias, interaction dynamics (following behavior), and network statistics.

Using bias labels sourced from the AllSides website, we compile a dataset encompassing six facets of news bias. By manually extracting follower/following relationships, we scrutinize interactions among 65 active Twitter news accounts. Despite the relatively small size of the network node data utilized, the results unveil discernible clustering patterns suggestive of echo chambers, with limited engagement between divergent ideologies. This study emphasizes the potential impact of bias on information dissemination and democratic expression. These findings provide valuable insights into the dynamics of echo chambers in modern social media landscapes and also advocate for enhanced regulation to promote diversification, particularly within news media.

## 2. Related Work

This study initially refers to a 2005 article [10] which employed a methodology wherein newspaper articles were tokenized, and a regression model mapping was conducted with the corresponding political party affiliation of the congressperson authors. Word scores were assigned to 1000 politically distinctive phrases, selected through chi-square analysis, and these word scores were then used to predict political bias in the remaining testing sample articles. The fundamental assumption of this methodology is that authors with different political inclinations will more frequently use specific vocabulary and phrases. Consequently, the overall bias of a media outlet can be further predicted through these politically colored terms.

---
[1] https://www.thesocialdilemma.com/

It exhibited a significant correlation (up to 0.4 with a p-value of 0.01) with bias classifications made by a website providing human judgment (Mondo Times) at that time.

Despite numerous investigations have employed both quantitative and qualitative methodologies to explore the hypothesis surrounding the existence and impact of echo chambers on social media, research in this field is inherently challenging and often relies on variables and data that are difficult to collect, measure, and interpret. A systematic review of literature [8] indicates a substantial influence of research methods and data sources on the derived conclusions. Quantitative studies generally affirm the existence and impact of echo chambers, whereas qualitative inquiries often challenge their presence and influence.

The fundamental reason may reside in the intrinsic drawbacks and limitations of methodological and conceptual choices. For instance, a media-centric focus on communication and interactions, especially relying on *quantitative* methods such as digital trace data of specific interaction networks, may introduce bias and disproportionately highlight the influence of a minority of users engaging in online interactions [8]. According to a survey, only 50% users actively participate in online interactions, such as reposts and comments, while a mere 20% hardly ever engage [11]. On the other hand, user-centric approaches centered around information exposure, often based on *qualitative* methods including small samples and retrospective self-reports, might yield inaccurate and biased outcomes and thus underestimate fragmentation and polarisation on social media [8].

Despite these challenges, combining self-reported data with digital trace data holds significant potential for future studies. An integrated approach could potentially offer a more comprehensive understanding of users' information environments by combining rich individual-level data with direct observations of online behavior in its natural context.

## 3. Research Questions

This article primarily investigates two questions. First, how to define and measure the interaction behavior between news media on social media platforms. Second, whether there is evidence that shows existence of echo chambers in Twitter's news ecosystem.

**RQ1**: How to measure Twitter news accounts' interactions?

**RQ2**: Whether echo chamber exist in Twitter composed of these news accounts?

## 4. Methodology and Data

### 4.1 Rating labels for news bias

This study utilized the website, AllSides[2], which provides human judgment, to furnish bias labels genuinely utilized in the data network research. This website rates the biases of over 460 news media outlets through two data sources: Blind Bias Survey and Editorial Review. By matching the names of these 460 media outlets with those in the 2005 study and conducting a bias relevance test, the correlation coefficient reached 0.454, with a p-value of 0.004767, indicating a highly significant correlation. Through cross-validation, the methodology employed in the 2005 study demonstrates robustness. The ratings from the AllSides website also exhibit validity and credibility, with a larger dataset and more comprehensive ratings across six dimensions: left, left-center, center, right-center, right, and allsides. In this study, Python was used to crawl the IDs and corresponding biases of over 460 news media outlets, creating a CSV file. This file serves as a sample space and bias labels for subsequent Twitter data mining and network analysis. In the later sections of this paper, we will represent the six dimensions of news bias rating—left, left-center, center, right-center, right, and allsides—with light blue, blue, yellow, light red, red, and green, respectively, facilitating visual presentation and intuitive understanding.

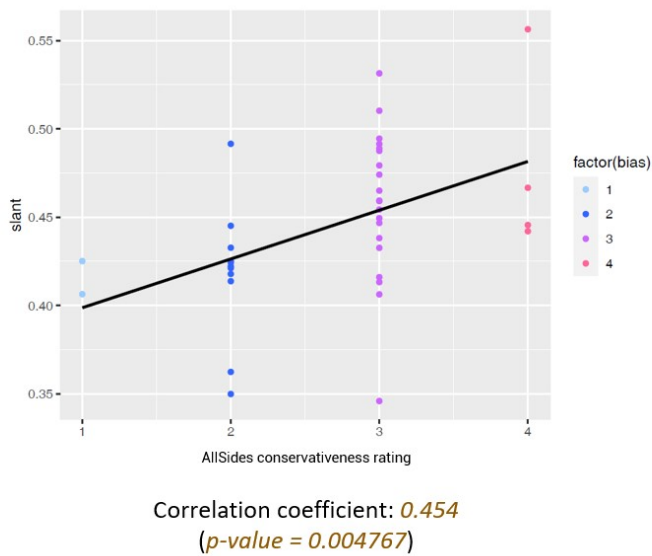 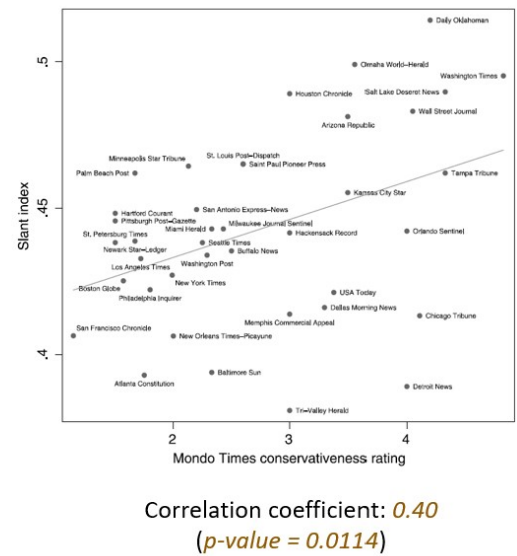

Figure 1: Correlation testing between ratings generated by text modeling and human judgment

---

[2] www.allsides.com/media-bias

Source: www.allsides.com/media-bias

Figure 2: Rating Labeling provided by the AllSides website

It is crucial to emphasize that within the AllSides website, the "allsides" label carries a significantly different implication from the "center" label. Although conventionally, one might assume that "center" denotes objectivity and neutrality, according to the AllSides website's definition, "center" merely signifies that these news media outlets consciously avoid using politically colored vocabulary or publishing articles with explicit political stances to prevent conflicts and maintain a neutral tone and ambiguous attitude, making it challenging to predict their true positions. On the other hand, the "allsides" label indicates that these media outlets deliberately present articles from different perspectives, fostering a scene of diverse communication and vigorous debate to broaden readers' perspectives and stimulate critical thinking. In the subsequent sections, we will combine statistical analysis to delve deeper into the examination of news media under these two labels.

### 4.2 Twitter news accounts' following/follower relationships

This study focuses on utilizing the follower/following relationship as a metric to measure interactions between Twitter news accounts. Due to limitations on free developer API access following Twitter's transformation into "X" by Elon Musk, the study could not employ automated crawling and mining of follower/following lists using the API. Fortunately, X provides an automated list called "Followers you know" on the profile pages of the accounts you follow, facilitating rapid list screening. Manual extraction and entry into an Excel matrix were then performed. From the over 460 news media outlets listed on AllSides, 65 active

Twitter news accounts with official IDs were selected, where activity was defined by a combined follower and following count (degree) exceeding 5. In Excel, a 65×65 matrix was constructed, with a value of 1 indicating that the ID in the row follows the ID in the column, and a value of 0 indicating no mutual following.

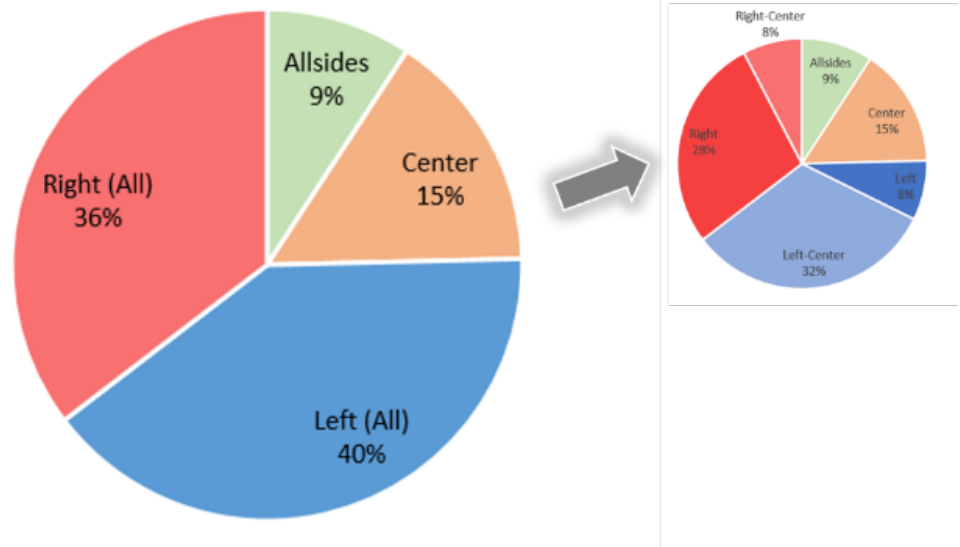

Figure 3: The composition of the six labels among the 65 news accounts

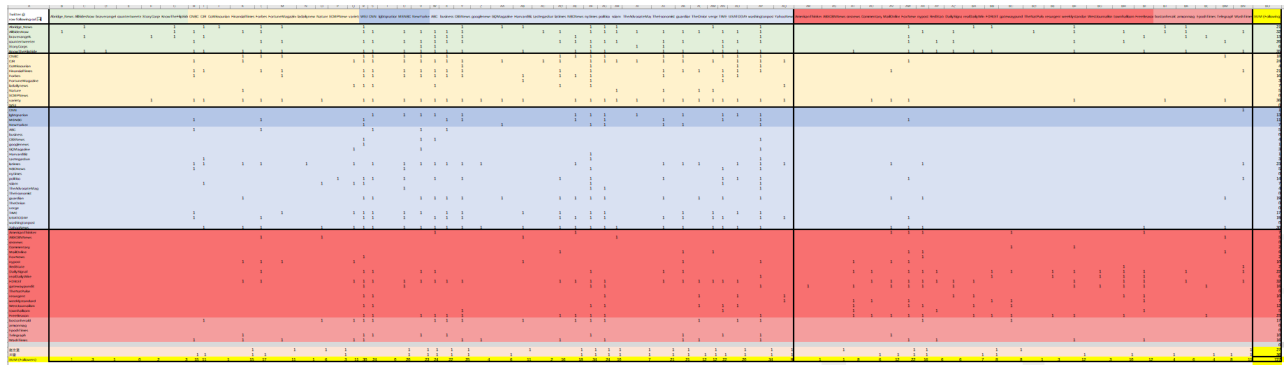

Figure 4: The following/follower matrix in Excel (colored by the row's news bias)

In summary, these two files (one bias labeling.csv and one following/follower matrix.xlsx) constitute the database for the data analysis in this study. The next section will employ Excel and Gephi for data processing and analysis.

## 5. Results and Analysis

### 5.1 Descriptive Statistics

As shown in Figure 5, listing all IDs with a following count exceeding 10, the accounts with the highest number of followers are primarily distributed among the allsides, center, and left-center factions. This aligns with the logical expectation that these three label factions exhibit a more inclusive and open-minded attitude. Particularly noteworthy is the ID variety, which follows a total of 36 other news accounts, constituting 56.25% of the overall interactions—a genuine representation of variety. It is worth noting that within the conservative faction, Federalist (FDRLST), FreeBeacon, and DailySignal also demonstrate a relatively diverse range of followed accounts.

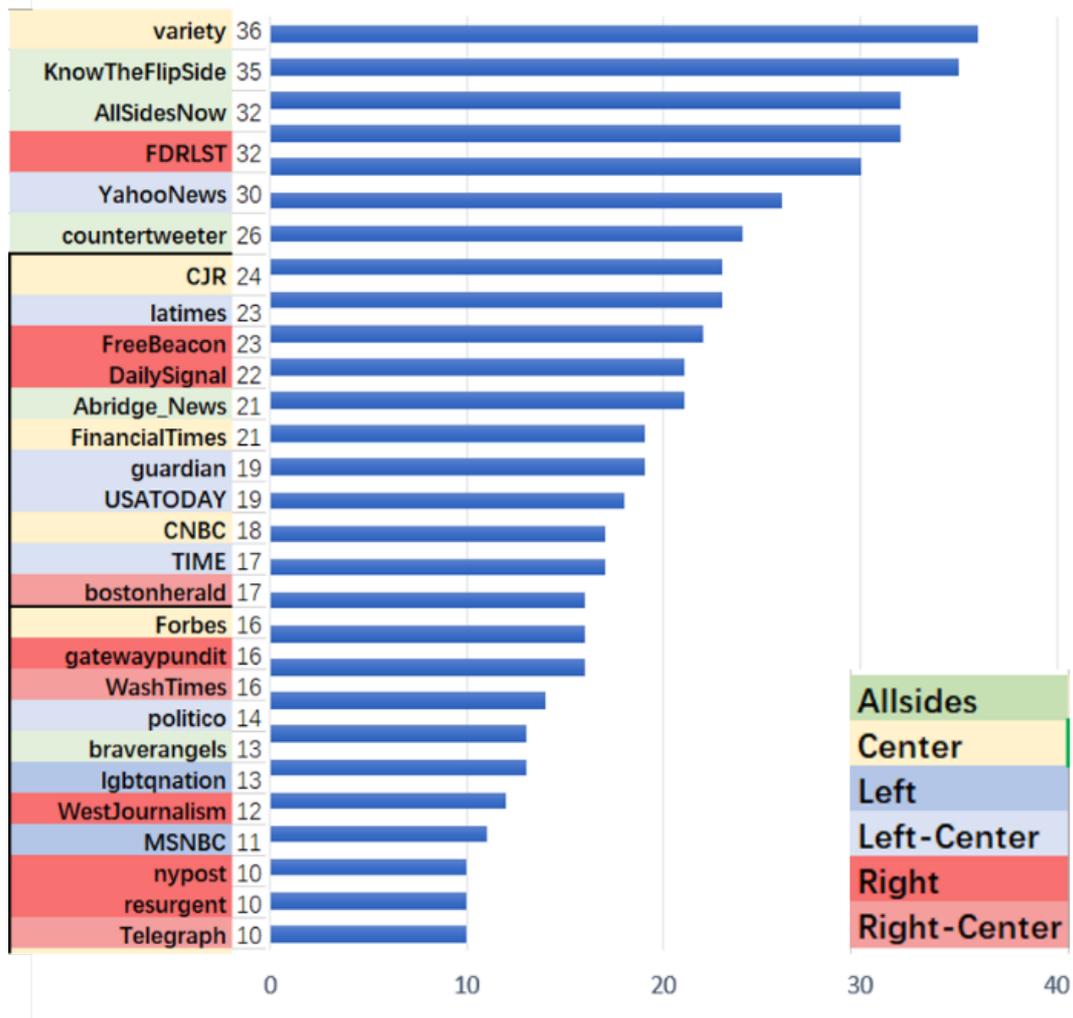

Figure 5: IDs with a following count exceeding 10

However, the overall following count may not entirely reflect diversity. After all, if an account follows only one faction, although the absolute count is high, the diversity is low. Therefore, in Table 1, we break down the data into specific categories of interest. The numerical value in each cell of Table 1 represents the following coefficient (F) for all news accounts of the corresponding bias type in the row towards all accounts of the same bias type in the column. The numerator of F is the actual number of followings, and the denominator is the theoretical maximum number of followings. Thus, the F value can indicate the following preference of a news account with a certain bias type (Following Preference).

The values in the diagonal of the table represent the F values for following accounts of the same type and can be considered as baselines (distributed between 0.2 to 0.25). F values higher than this range indicate excessive following, while values lower than this range indicate insufficient following. Therefore, based on the table, it is evident that, except for the allsides type, which is a niche account with an overall low following count, the other cells reflect clear preference types. For instance, AllSides and Center, in reality, follow the left-wing more (0.47 and 0.40). AllSides follows the right-wing at a level consistent with the baseline, but Center follows the right-wing significantly too little (0.04). Left-wing follows the Center and right-wing significantly too little (0.04 and 0.02). In comparison to the purely left-wing, left-center follows the Center significantly more (0.12 > 0.04) and the right-wing slightly more (0.03 > 0.02). Surprisingly, the right-wing, known for being conservative, follows dissenting Center and left-wing accounts more than expected, and even the F value for right-center exhibits a more left-leaning tendency. This situation may be due to some accounts in the right and right-center factions having broader perspectives and thinking, or it could be a result of obtaining intelligence by directly following the opposing faction. It may also be influenced by a bias in the labeling of the right-wing faction.

|  | Allsides | Center | Left | Right |  | null |
|---|---|---|---|---|---|---|
| Allsides | 0.25 | 0.27 | 0.47 | 0.25 |  |  |
| Center | 0.02 | 0.20 | 0.40 | 0.04 |  | Baseline |
| Left | 0.00 | 0.04 | 0.22 | 0.02 |  |  |
| Left-Center | 0.00 | 0.12 | 0.21 | 0.03 |  | Unbalanced |
| Right | 0.00 | 0.10 | 0.15 | 0.21 |  |  |
| Right-Center | 0.00 | 0.16 | 0.22 | 0.06 |  | Out-of-Box |

Table 1: Following Preferences of six factions

In conclusion, through the above analysis, the overall following behavior and preferences of each faction align with logical expectations. Instances where a few following preferences deviate from the anticipated norms may present multiple plausible explanations and merit further data collection for in-depth investigation.

**5.2 Following Network Analysis**

In this section, we will explore the third and final research question, which is the central focus of this paper: Are There Echo Chambers in the US News Ecosystem? By importing the CSV and Excel files into Gephi, we obtained typical Network Statistics as presented in Table 2. Based on the provided statistical results, the network can be interpreted and described as follows:

- The **average degree** is 10.215, indicating that each node is, on average, directly connected to 10.215 other nodes. This suggests a relatively high level of connectivity among nodes in the network.
- The **density** is 0.16, representing the ratio of actual edges to possible edges in the network. This value is relatively low, indicating sparse connections among nodes in the network.
- The **diameter** is 6, signifying that the maximum length of the shortest path between any two nodes in the network is 6. This means that any two nodes can be connected through a network within at most 6 intermediate nodes, to some extent validating the theory of six degrees of separation.
- The **average path length** is 2.277, representing the average shortest path length between any two nodes in the network. This indicates a relatively close interconnection among nodes in the network.
- The **clustering coefficient** is 0.3, signifying the degree of clustering among nodes in the network. This relatively high value suggests the presence of clustering phenomena in the network, meaning nodes tend to form clusters with neighboring nodes.
- The **modularity** is 0.16, indicating the existence of 3 communities or modules in the network. Modularity measures the strength of community structure in the network, indicating that nodes can be divided into 3 independent groups based on some criteria.

Based on these statistical results, the network can be inferred to possess the following characteristics:

The network exhibits significant connectivity, fostering interaction and information dissemination. Despite these connections, the network's density is low, indicating sparse interconnections. Short average shortest paths between nodes suggest rapid information spread. Clustering effects in the network show nodes forming groups, aiding information dissemination within clusters but limiting propagation between them. With 3 communities matching the primary labels (right, center, left), the network is somewhat divided into independent groups, fostering concentrated information dissemination within each, creating a degree of "clustering effect" or "echo chamber effect."

Figure 6: Three Gephi-generated communities matching the primary labels (right, center, left)

| Network Statistics | |
|---|---|
| # Node | 65 |
| # Edge (directed) | 664 |
| Average Degree | 10.215 |
| Density | 0.16 |
| Diameter | 6 |
| Average Path Length | 2.277 |
| Cluster Coefficient | 0.3 |
| Moduality | 0.16 |
| Number of communities | 3 |

Table 2: Network Statistics generated by Gephi

By further leveraging the visualization capabilities of Gephi, we obtained the two Network Visualizations below, where colors represent the bias labels assigned earlier. The left image is a simplified graph produced by certain parameters, while the right image is an advanced graph generated through the Fruchterman Reingold algorithm. This algorithm simulates physical particle properties (nodes exert repulsion, and edges provide attraction, achieving balance by minimizing total energy), offering a more vivid and intuitive representation of the interaction dynamics between different nodes.

Overall, the network graphs with label-assisted coloring visually depict the clustering patterns among different factions. The distribution of nodes in red and blue colors generally indicates the opposition between left-wing and right-wing factions. Nodes representing

AllSides in green and Center in yellow are more concentrated in the left-wing faction, aligning with the observations from the F values in the preceding section on following preferences. Certain nodes representing right-wing in red and right-center in light red deviate further from the cluster of red nodes, accurately reflecting the situation of conservative accounts with significant followings, such as Federalist (FDRLST), FreeBeacon, and DailySignal.

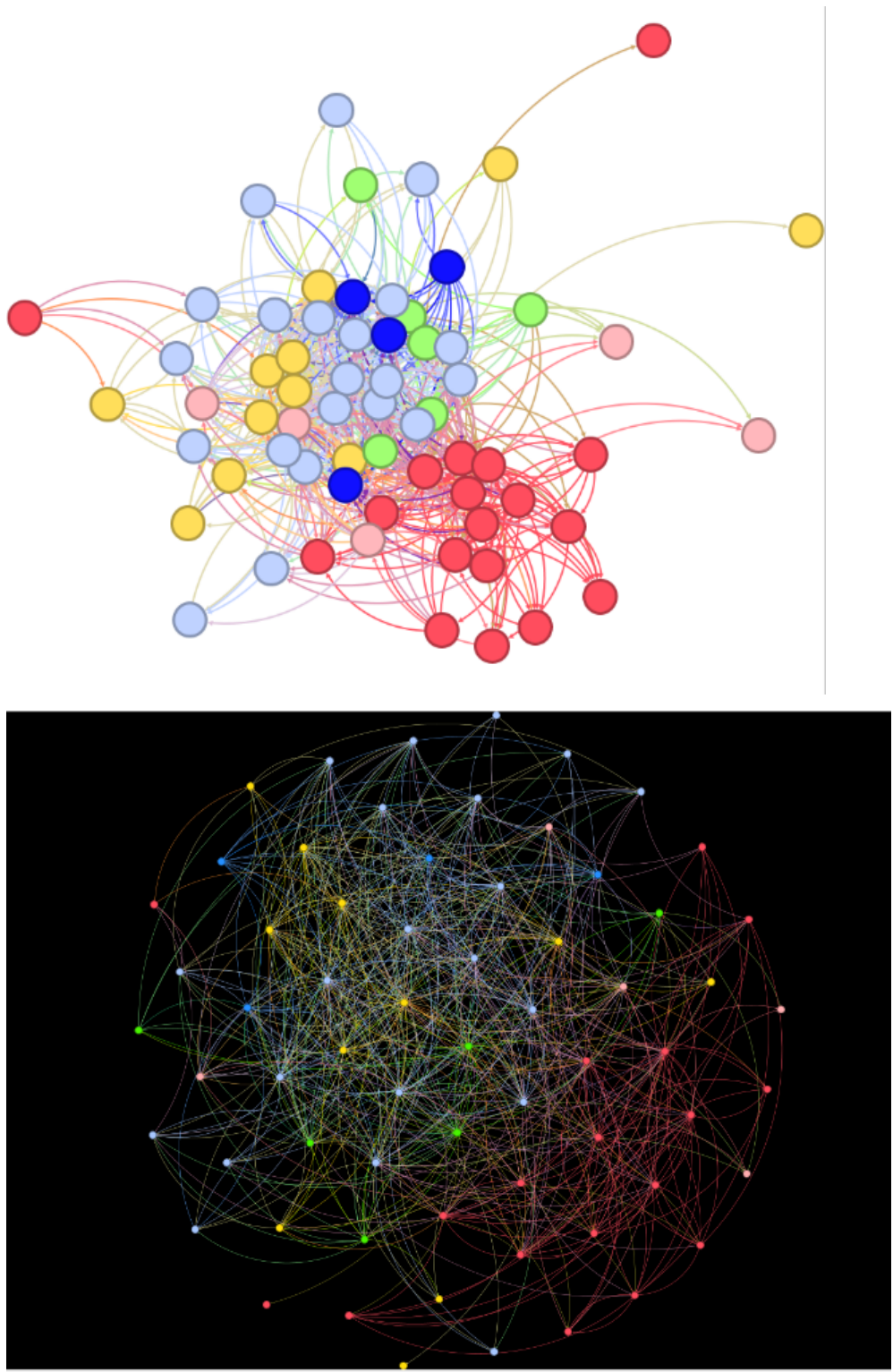

Figure 7: Network Visualisation generated by Gephi

## 6. Conclusion

In conclusion, although the sample size in this study is relatively small and only considers the dynamic interactions among official news accounts, without incorporating data analysis for reader nodes, the network statistics and visualization in this section provide some degree of quantitative and visual evidence for the existence of an echo chamber effect. Thus, we conclude that in this network comprising 65 nodes with different political inclinations in news accounts, an echo chamber effect does indeed exist. This, to some extent, may impact the transmission of diverse information and the expression of democratic politics.

## 7. Limitations and Future Directions

As mentioned above, a significant limitation of this study is the small sample size and the lack of sampling for audience and other interaction types (Need to explore beyond the following/follower relationship). In the future, obtaining multiple recursive lists of followings and interactions (posts, reposts, likes, comments, etc.) with higher-level API permissions will yield more comprehensive, accurate, and scientifically analyzed network interactions.

Additionally, for bias measurement in RQ1, future research can adopt more advanced tools to validate bias labels (e.g., NLP, GPT). Moreover, the following behavior of official accounts itself also exhibits bias. For instance, official institutions often have stricter operational and management mechanisms, leading to their following behavior being potentially linked to business logic or management strategies. This reflects a form of strategic selective exposure, etc. Therefore, it may not accurately reflect their true following preferences.

Finally, the network lacks weighted edges reflecting different types of followings, such as scoring fewer points for following similar accounts and more points for following dissimilar accounts. Weighted edges of this nature could yield different results in network statistics and visualization, thereby affecting the analysis of the existence of an echo chamber.